\documentclass[aps,prl,twocolumn,groupedaddress,superscriptaddress,amsfonts,
amssymb,amsmath,citeautoscript,a4paper,english]{revtex4-1}
\usepackage[T1]{fontenc}
\usepackage[latin9]{inputenc}
\usepackage{amsmath}
\usepackage{amssymb}
\usepackage{graphicx}
\usepackage{xcolor}

\usepackage{babel}
\begin{document}
\title{Quantum surface effects in strong coupling dynamics}
\author{Vasilios Karanikolas}
\email{KARANIKOLAS.Vasileios@nims.go.jp}
\affiliation{International Center for Young Scientists (ICYS), National Institute
for Materials Science (NIMS) 1-1 Namiki, Tsukuba, Ibaraki 305-0044, Japan}
\author{Ioannis Thanopulos}
\affiliation{Materials Science Department, School of Natural Sciences,
University of Patras, Patras 265 04, Greece}
\author{Joel~D. Cox}
\affiliation{Center for Nano Optics, University of Southern Denmark, Campusvej 55,
DK-5230 Odense M, Denmark}
\affiliation{Danish Institute for Advanced Study, University of Southern Denmark,
Campusvej 55, DK-5230 Odense M, Denmark}
\author{Takashi Kuroda}
\affiliation{National Institute for Materials Science (NIMS) 1-1 Namiki, Tsukuba,
Ibaraki 305-0044, Japan}
\author{Jun-ichi~Inoue}
\affiliation{National Institute for Materials Science (NIMS) 1-1 Namiki, Tsukuba,
Ibaraki 305-0044, Japan}
\author{N.~Asger Mortensen}
\affiliation{Center for Nano Optics, University of Southern Denmark, Campusvej 55,
DK-5230 Odense M, Denmark}
\affiliation{Danish Institute for Advanced Study, University of Southern Denmark,
Campusvej 55, DK-5230 Odense M, Denmark}
\author{Emmanuel Paspalakis}
\affiliation{Materials Science Department, School of Natural Sciences,
University of Patras, Patras 265 04, Greece}
\author{Christos~Tserkezis}
\email{ct@mci.sdu.dk}
\affiliation{Center for Nano Optics, University of Southern Denmark, Campusvej 55,
DK-5230 Odense M, Denmark}

\date{\today}
\begin{abstract}
Plasmons in nanostructured metals are widely utilized to trigger
strong light--matter interactions with quantum light sources. While the nonclassical
behavior of such quantum emitters (QEs) is well-understood in this context, the role
of quantum and surface effects in the plasmonic resonator is usually neglected. Here,
we combine the Green's tensor approach with the Feibelman $d$-parameter formalism to
theoretically explore the influence of quantum surface effects in metal-dielectric
layered nanostructures on the relaxation dynamics of a proximal two-level QE. Having
identified electron spill-out as the dominant source of
quantum effects in jellium-like metals, we focus our study on sodium. Our results
reveal a clear splitting in the emission spectrum, indicative of having reached the
strong-coupling regime, and, more importantly, non-Markovian relaxation dynamics of
the emitter. Our findings establish that strong light--matter coupling is not suppressed
by the emergence of nonclassical surface effects in the optical response of the metal.
\end{abstract}
\maketitle

{\it Introduction} --- Confining light in sub-diffraction volumes via surface
plasmon polaritons (SPPs) has become common practice to enhance nanoscale light--matter
interactions in the past decade~\cite{ozbay_sci311}. In combination with multi-level
quantum emitters (QEs), plasmonic nanostructures play the role of the
cavity~\cite{torma_rpp78,Pelton_natphot,bozhevolnyi_natphot2017,baumberg_natmat2019}
in cavity quantum electrodynamics (QED)~\cite{Mandel_Cambridge1995,Scully_Cambridge1997},
while the QEs, depending on the desired functionality, can be natural (atoms, molecules)
or artificial (quantum wells and dots, defects in nanodiamonds, collective states in
transition metal dichalcogenides or hexagonal boron nitride)~\cite{akimov_nat450,
chang_natphys3,goban_natcom5,hood_pnas113,fernandez_acsphot5,koperski_pnas117}.
Advanced fabrication techniques now enable the engineering of metal cavities with
characteristic dimensions below $10$\;nm, as well as precise positioning of QEs within them,
to produce promising templates for bright single-photon emitters~\cite{manolatou_ieeejqe44,
akselrod_natphot8,rose_nl14,chikkaraddy_nat535,May2020}. The plasmonic cavity is typically formed
by a metallic substrate interacting with another metallic film or with dropcasted nanoparticles;
separating the two plasmonic components by a thin dielectric spacer containing QEs in
the form of defects, molecules, or quantum dots~\cite{bogdanov_nl18,katzen_acsami12,
ojambati_acsphot7,Ding_nanophotonics}, the resulting intense electromagnetic (EM) fields,
confined within minimized volumes, trigger extreme light--matter interactions.

Traditionally, in the local-response approximation (LRA)~\cite{Hohenester_Springer2020},
the optical response of metals is described through the Drude model, or through
experimentally measured bulk permittivities. But, as plasmonic cavities become
narrower, the interaction strengths with the therein confined QEs are overestimated
by LRA, as compared to experimental results~\cite{ciraci_sci337,savage_nat491,
scholl_nat483}, calling for amendment of the theoretical models~\cite{romero_oex14,
esteban_natcom3,luo_prl111,raza_nanophot2,yang_nat576}; the main missing element is
information about quantum effects in the metal. As a first extension of LRA, the
hydrodynamic Drude model (HDM), which describes the motion of the compressible
free-electron gas as a convective fluid, has met with considerable success~\cite{moreau_prb87,
raza_prb88,mortensen_natcom5,tserkezis_acsphot5b}, yet effects associated with
electron spill-out still require a self-consistent treatment~\cite{toscano_natcom6,
yan_prb2015,ciraci_prb93,ciraci2019}. However, the ultimate goal of an \emph{ab initio}
optical response calculation for structures with sizes of order $100$\;nm at room
temperature remains out of computational reach~\cite{varas_nanophot5}.

A more efficient approach is based on the work of Feibelman~\cite{feibelman_pss12},
who bridged EM and \emph{ab initio} calculations through appropriate surface-response
functions, the $d$-parameters ($d_{\perp}$ and $d_{\parallel}$), which need be calculated only once for a given surface
to be implemented in a quantum-informed surface-response formalism
(SRF)~\cite{yan_PRL2015,christensen_prl118,yang_nat576,Goncalves_natcom11,Mortensen2021}.
The full QED problem is then described by quantum-corrected mesoscopic boundary conditions
at the metal surface~\cite{yan_PRL2015,yang_nat576}. Macroscopic QED~\cite{Rivera2020,
Head-Marsden2020} allows to directly apply the rigorous Green's function approach
developed within LRA, now corrected by SRF, without invoking quantum effects in a
quasi-normal mode formalism~\cite{Lalanne_lpr2018,Dezfouli_optica2017,Franke_PRL2019}.

The relaxation process of a QE placed in a nanostructured environment has been
extensively investigated in the context of the macroscopic QED theory. The photonic
environment provides
an enhanced density of optical states ---often, but not restrictively, in the
form of an optical resonance---
at a specific energy that, when matched with the transition energy of the QE,
leads to considerable QE--environment light--matter coupling-strength enhancement.
Usually, the interaction is described within a Lorentzian model of the spectral
density which, under strong-coupling conditions, results in a Rabi splitting in the
emission spectrum and oscillations in the relaxation dynamics of the QE. However,
this approach fails to describe possible partially excited state population trapping
effects~\cite{Yang2017,thanopulos_prb99,Yang2019}. Here, we go beyond this Lorentzian
spectral density model approach, taking into account the influence of the exact
QE--nanostructure spectral density on the light--matter interaction strength.

To date, focus has been mostly cast on the interaction of QEs with plasmonic
nanostructures described by nonlocal dielectric functions within HDM, identifying
significant differences in both weak~\cite{tserkezis_nscale8,christensen_nn8} and
strong-coupling regimes~\cite{tserkezis_acsphot5a,tserkezis_rpp83}, as compared to
LRA. Here, we go one step further to explore, for the first time, the role of
quantum-informed plasmonics in the population dynamics of the QE excited state
under strong-coupling conditions, where QE and environment coherently exchange
energy. We focus on insulator/metal (IM), insulator/metal/insulator (IMI), and
metal/insulator/metal (MIM) geometries~\cite{smith_nanoscale,raza_prb88}, for
which the strong-coupling regime proves reachable even when quantum corrections
in the metal are taken into account.

{\it Theory} --- The QE is approximated as a two-level system with ground state
$|g\rangle$ and excited state $|e\rangle$, transition frequency $\omega_{0}$,
and dipole moment $\boldsymbol{\mu}$. Initially, the QE is in the excited state,
and the EM field is in its vacuum state $|i\rangle=|e\rangle\otimes|0\rangle$,
while the final state is $|f\rangle = |g\rangle \otimes
\hat{f}_{i}^{\dagger}(\mathbf{r}, \omega)|0 \rangle$, where the QE relaxes to
the ground state by emitting a photon or exciting SPPs dressed states supported by
the metallic nanostructures~\cite{dung_pra65}, and $\hat{f}^{\dagger}$ denotes the
bosonic creation operator of the dressed state $i$. The QE relaxation rate $\Gamma$
is given by $\Gamma (\mathbf{r}, \omega) = \Gamma_{0} \tilde{\Gamma}(\mathbf{r}, \omega)$,
where $\Gamma_{0} = \omega^{3} \mu^{2}/3 \pi c^{3} \hbar \varepsilon_{0}$ is the
vacuum rate and $\tilde{\Gamma} (\mathbf{r}, \omega)$ is the Purcell factor of a QE
placed at $\mathbf{r}=(0,0,z)$ above a metal/dielectric interface; for a transition
dipole moment perpendicular to a single insulator/metal interface, the Purcell factor
has the form
\begin{equation}\label{eq:02}
\tilde{\Gamma}_{z} (\mathbf{r}, \omega) = 
\sqrt{\varepsilon_{d}} - \frac{3c}{2\omega}
\mathrm{Re} \Bigg\{\int\limits_{0}^{\infty} \text{d} k_{s}
\frac{k_{s}^{3}R_{\rm TM}}{k_{d,z}k_{d}^{2}}
e^{2i k_{d,z}z}\Bigg\},
\end{equation}
where $R_{\rm TM}$ is the generalized transverse magnetic (TM)
Fresnel coefficient~\cite{Tai_Oxford1994,Scheel2009}.
We note that the metal-dielectric structures considered
here are as close to QED cavities as possible, dominated by a single mode,
thus allowing the use of a Purcell factor without ambiguity~\cite{koenderink_ol35}.

Our material of choice is sodium (Na), whose work function is lower than that
of the common plasmonic metals gold and silver, so as to focus on electron
spill-out effects~\cite{abajo_jpcc112,echarri_arxiv2020}. For the single
dielectric/metal interface, the TM reflection coefficients take the form
(assuming $d_\parallel=0$, valid for Na)~\cite{Goncalves_natcom11}
\begin{equation}\label{eq:03a}
R_{\rm TM} =
\frac{\varepsilon_{\rm m} k_{{\rm d},z} - \varepsilon_{d}k_{{\rm m},z} +
\mathrm{i} \left(\varepsilon_{\rm m}-\varepsilon_{\rm d}\right) k_{s}^{2}
d_{\perp}(\omega)}
{\varepsilon_{\rm m} k_{{\rm d},z} +
\varepsilon_{\rm d} k_{{\rm m},z} -
\mathrm{i} \left(\varepsilon_{\rm m} - \varepsilon_{\rm d}\right)
k_{s}^{2}d_{\perp}(\omega)},
\end{equation}
where $\varepsilon_{\rm d}$ and $\varepsilon_{\rm m}$ denote the permittivity
of the dielectric ($\mathrm{d}$) and metal ($\mathrm{m}$), respectively,
$k_{0} = \omega/c$ is the free-space wave vector, $k_{j} =\sqrt{\varepsilon_{j}} k_0$
(with $j = \mathrm{m,d}$) is the wavevector of each medium, analyzed in in-plane
($k_\mathrm{s}$) and normal ($k_{j,z} = \sqrt{k_{j}{2} - k_{\mathrm{s}}^{2}}$) components.
For Na, we use a Drude model $\varepsilon(\omega) = 1 - \omega_{\rm p}^{2}/
\left(\omega^{2} + \mathrm{i} \omega \gamma\right)$, with $\omega_{\rm p}$ being the
plasma frequency and $\gamma$ the damping rate, taken as $\hbar \omega_{\rm p}= 5.9$\;eV
and $\hbar \gamma=0.1$\;eV, while $d_{\perp}$ is given by Lorentzian-fitted data extracted
from \emph{ab initio} calculations for jellium~\cite{christensen_prl118,Goncalves_natcom11}
(see Supplemental Material~\cite{SupportingInformation}).

The relaxation of the quantum emitter is described by its emission
spectrum~\cite{vanvlack_prb85,Karanikolas2020}
\begin{equation}\label{eq:05}
S (\omega, \mathbf{r}) = 
\frac{1}{2\pi} 
\left| \frac{(\mu^{2} \omega^{2})(\varepsilon_{0} c^{2})
\mathbf{\hat{n}} \cdot \mathbf{G} (\omega,\mathbf{r}, \mathbf{r}_{d})}
{\omega_{0} - \omega- \int_{0}^{\infty} \mathrm{d} \omega^{\prime}
J(\omega_{0}, \omega^{\prime}, \mathbf{r})/(\omega^{\prime} - \omega)}\right|^{2}
\end{equation}
in the frequency domain, and by the time-dependent QE excited state probability
amplitude, $c_{1}(t)$, given by the solution of the integro-differential
equation~\cite{Tudela2014a,Li2016PRLa,thanopulos_prb95,thanopulos_prb99}
\begin{equation}\label{eq:04}
\frac{\mathrm{d} c_{1}(t)}{\mathrm{d}t}
=\mathrm{i} \int_{0}^{t} \mathrm{d} t^{\prime} K(t-t^{\prime})
c_{1}(t^{\prime}).
\end{equation}
The kernel of Eq.~(\ref{eq:04}) is given by $K(\tau) = \mathrm{i}
e^{\mathrm{i} \omega_{0}\tau} \int_{0}^{\infty} \mathrm{d} \omega
J(\omega_{0}, \omega, \mathbf{r}) e^{-\mathrm{i}\omega\tau}$, where
$J(\omega_{0}, \omega, \mathbf{r}) = \tilde{\Gamma}_{z} (\omega, \mathbf{r})/
[2\pi\tau_{0}]$ is the spectral density, $\omega_{0}$ is the energy difference
between the ground and excited QE states, and $\tau_{0}= 1/\Gamma_{0}(\omega_{0})$ is
its free-space lifetime~\cite{Tudela2014a,Li2016PRLa,thanopulos_prb95,thanopulos_prb99}.
In Eq.~(\ref{eq:05}), $\mathbf{r}_{d}$ is the position where the signal is detected,
and $\mathbf{r}$ is the position of the QE, while $\omega$ is the emission frequency of
the combined QE--Na system~\cite{vanvlack_prb85,Karanikolas2020}. Eq.~(\ref{eq:05}) implies
that, if the QE--nanostructure coupling strength is high enough, the emission spectrum
will feature a doublet of emission peaks, whose energy difference defines the Rabi
splitting $\hbar\Omega$. More details about the macroscopic QED model used can be found
in the Supplemental Material~\cite{SupportingInformation}.
Solving Eq.~(\ref{eq:04}), the population dynamics of the QE is calculated
under the rotating wave approximation (RWA), which is well-justified under the conditions
considered here; as we show in~\cite{SupportingInformation}, results obtained by relaxing
the RWA exhibit the same quantitative behavior~\cite{thanopulos_prb95,thanopulos_prb99}.

\begin{figure}[h]
\centering
\includegraphics[width=0.5\textwidth]{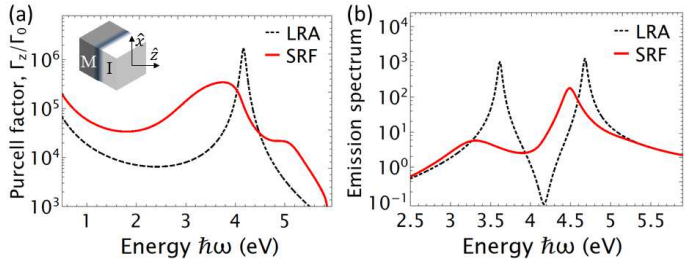}
\caption{A QE with $z$-oriented transition dipole moment, placed $1\,$nm away
from an air/Na IM interface. (a) Purcell factor and (b) emission spectrum
(arbitrary units)
within LRA (black dashed lines) and SRF (red solid lines).}\label{fig:01}
\end{figure}

{\it Results and discussion} --- 
To investigate the role of spill-out on the emission properties of QE, we first revisit
the simple IM case involving a single SPP resonance. We note here that, in general,
there is a competition between relevant contributions from SPPs and a pseudomode
stemming from EM modes with higher in-plane momentum~\cite{Tudela2014a,delga_prl112};
effectively, which contribution dominates depends on the QE--metal separation. In
Fig.~\ref{fig:01}(a) we present the Purcell factor as a function of the QE transition
energy, for a QE  placed $1\,$nm away from an air--Na IM interface, as evaluated within
LRA and SRF. We  directly observe that the highest Purcell factor value within SRF has a 
redshifted peak value that is one order of magnitude smaller than the corresponding
LRA result, and the SPP-originated resonance around $\hbar\omega =
\hbar\omega_{\rm p}/\sqrt{2} = 4.125$\,eV becomes significantly broader. These
alterations are related to additional material losses due to surface-enabled Landau
damping~\cite{tserkezis_prb96}. Nevertheless, at lower frequencies, away from the SPP 
resonance, the SRF Purcell factor is in fact higher than the one predicted by LRA.
Similar results have been discussed in Ref.~\cite{Goncalves_natcom11}, thus validating
the Green's tensor formalism employed here.
Interestingly, a second spectral feature appears at around $5$\;eV
as a shoulder in the Purcell spectrum; this is directly related to spill-out, and
is known as the Bennett multipole surface plasmon~\cite{bennett_prb1,svendsen_jpcm32}.

The emission spectrum of a QE with transition energy $\hbar\omega_{0} =
4.125\,$eV and free-space lifetime $\tau_{0}=0.1$\,ns is presented in
Fig.~\ref{fig:01}(b), where the Purcell factors of Fig.~\ref{fig:01}(a) are
used. In the LRA description, the emission spectrum features two sharp peaks,
splitting by $\hbar \Omega = 1.06$\,eV around the QE transition energy. When the
quantum aspects of the response of the metal are introduced through SRF, the two
emission peaks become broader, due to the higher material losses, but the Rabi
splitting increases to $\hbar \Omega = 1.3$\,eV; the latter phenomenon stems from
the dependence of the QE emission on the \emph{entire spectrum} of the Purcell factor,
which is higher within SRF away from the SPP resonance
[Fig.~\ref{fig:01}(a)]~\cite{christensen_prl118}.

\begin{figure}[h]
\centering\includegraphics[width=0.5\textwidth]{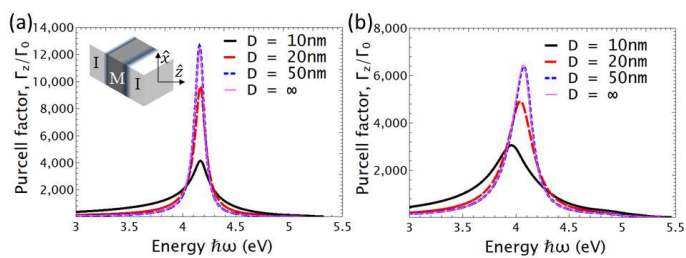}
\caption{Purcell factor within (a) LRA and (b) SRF of a QE placed $5$\,nm away from an IMI nanostructure, for different metal thicknesses $D = 10$ (black solid line), 20 (red dotted line) and $50$\;nm (blue dashed line). The thin purple shows the corresponding IM result.}\label{fig:02}
\end{figure}

Turning to the IMI case, where air is considered as the insulating medium, we
examine the corresponding Purcell factor in Figs.~\ref{fig:02}(a) and (b) within
LRA and SRF, respectively, for metal layer thicknesses $D=10,\,20$, and $50\,$nm.
Specifically, we set the distance between the QE and the insulator/metal interface
to $z_{\text{QE}}=5$\,nm, noting that the Purcell factors calculated for the IM and
the IMI geometry exhibit negligible differences for smaller values $z_{\text{QE}}
\sim 2\,$\,nm. In the case of LRA, the Purcell factor has its highest value at the
SPP energy for all layer thicknesses. On the other hand, when SRF is taken into account,
the maximum Purcell factor is roughly halved, and its peak positions redshift
significantly as the metal thickness decreases. For the thinnest layer ($D=10\,$nm),
the resonance within SRF has shifted by more than $\hbar \omega=0.2\,$eV compared to
LRA. For increasing metal thickness (above $D = 50\,$nm), the Purcell factor for the
IMI case converges to the IM result.

\begin{figure}[h]
\centering\includegraphics[width=0.5\textwidth]{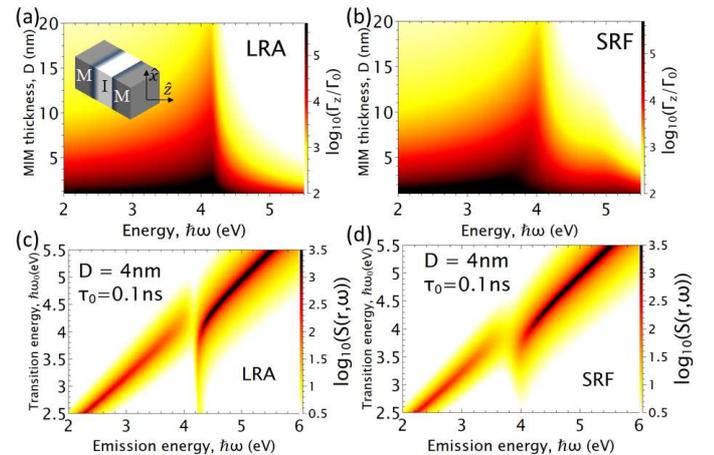}
\caption{Contour plot of the Purcell factor of a QE, as a function of emission
energy and dielectric layer thickness $D$, calculated within (a) LRA and (b) SRF,
and the corresponding contours of the emission spectra
(arbitrary units)
as a function of the QE emission and transition energies obtained with
(c) LRA and (d) SRF. The QE, with $\tau_{0} = 0.1\,$ns, is placed in the middle
of the insulator layer of an MIM structure with dielectric-layer thickness
$D = 4\,$nm.}\label{fig:03}
\end{figure}

While the IMI architecture can be instructive, in the remainder of the paper we
will focus on the MIM structure, which is expected to display many of the advantages
of plasmonic cavities discussed in the introduction ---apart, perhaps, from the antenna
effect that characterizes nanoparticle-on-mirror cavities~\cite{baumberg_natmat2019},
which contain the QE within them and, when relatively large nanoparticles are
involved, one can approximately assume that the QE experiences (locally) the cavity as
a MIM. The investigation of QE relaxation inside MIM structures is thus an important
first (exactly solvable) step towards understanding timely experimental works.

In Figs.~\ref{fig:03}(a,b) we compare LRA and SRF contour plots of the Purcell factor
as a function of the QE emission energy and the thickness $D$ of the insulator layer
(assumed air here) for a QE placed in the middle of the insulating gap
($z_{\text{QE}}=0$). In both plots, we observe that, for small $D$, the Purcell factor
is significantly enhanced compared to the reference free-space value. In LRA, the peak
of the Purcell factor coincides with the SPP resonance, while for higher energies there
is a sharp drop of its value, related to the absence of plasmon polariton modes in the
bandgap region of the MIM structure~\cite{marocico_pra84}. On the other hand, SRF reproduces
the expected redshift of the Purcell factor spectrum, while considerable enhancement is
observed inside the bandgap, as a result of the quantum effects captured by the Feibelman
parameters. As the thickness of the insulator increases, surface effects become less
important and the spectra obtained within the two models converge. What is evident from the
large Purcell-factor enhancements of Figs.~\ref{fig:03}(a,b) is that the QE--MIM interaction
can enter the strong coupling regime, as verified by the emission-spectrum contour plots
of Figs.~\ref{fig:03}(c,d), where clear anticrossings can be observed not only within LRA,
but also in the SRF case. Naturally, the emission double-peak features are sharper in the
LRA description, and are broadened when additional loss channels due to surface effects are
considered, but the observed Rabi splitting clearly survives.

\begin{figure}[h]
\centering\includegraphics[width=0.5\textwidth]{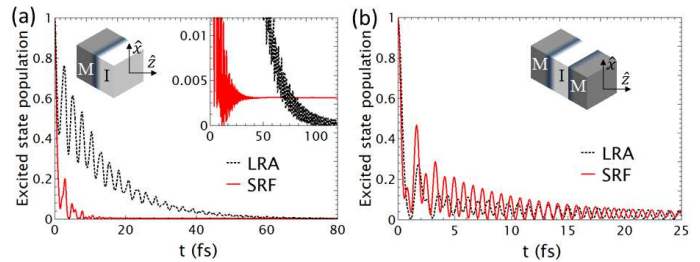}
\caption{Population of the excited state of a two-level QE with $\hbar \omega_{0}
= 3\,$eV and $\tau_{0} = 0.5\,$ns, as a function of time $t$. (a) The QE is placed
$1\,$nm away from an IM geometry. (b) The QE is placed in the middle of a MIM geometry
with dielectric-layer thickness $D = 2\,$nm. In both panels, black and red lines correspond
to results within LRA and SRF, respectively.}\label{fig:04}
\end{figure}

Having established the strong coupling regime, we turn now to the main focus of this paper,
which is the QE dynamics. The QE excited-state population dynamics presented in
Fig.~\ref{fig:04} exhibits strong oscillations, with very different features from
simple exponential relaxation within the Markovian (weak coupling) approximation,
where memory effects in Eq.~(\ref{eq:04}) are ignored. For a QE with transition energy
$\hbar \omega_{0} = 3\,$eV and vacuum relaxation rate $\tau_{0} = 0.5\,$ns,
Fig.~\ref{fig:04}(a) shows the excited state dynamics when $z_{\text{QE}} = 1\,$nm in
the IM geometry. In the LRA description, we observe that the relaxation follows an
exponential pattern modified by rapid oscillations, where after a few of
them the QE relaxes to the ground state within a time-span of $60\,$fs. Faster
relaxation to the ground state is predicted in SRF, although a closer look in the
inset of Fig.~\ref{fig:04}(a) reveals that in this case the excited state remains
partially populated, $\left| c_{\text{nL}} \right|^{2}=0.003$, even at longer times, although is expected to fully relax due to interaction with the environment.

To directly compare the IM and MIM geometries, we consider a QE centered in an
insulating (air) layer of $D = 2\,$nm thickness sandwiched between metal regions,
i.e., maintaining a $1\,$nm distance from the metal/dielectric interfaces. Here,
the QE population dynamics presented in Fig.~\ref{fig:04}(b) is characterized by
strong oscillatory behavior, with larger population values persisting over longer
time spans. While in LRA calculations the QE eventually relaxes 
to the ground state within the considered time scale, the SRF calculations accounting 
for the quantum response predict that $\sim 3$\% of the initial population 
remains in the excited state on the same time scale, a $10$-fold increase compared 
to the IM structure. The time scale of the plot is short compared 
to the QE lifetime, and complete relaxation will ultimately be achieved.

\begin{figure}[h]
\centering\includegraphics[width=0.5\textwidth]{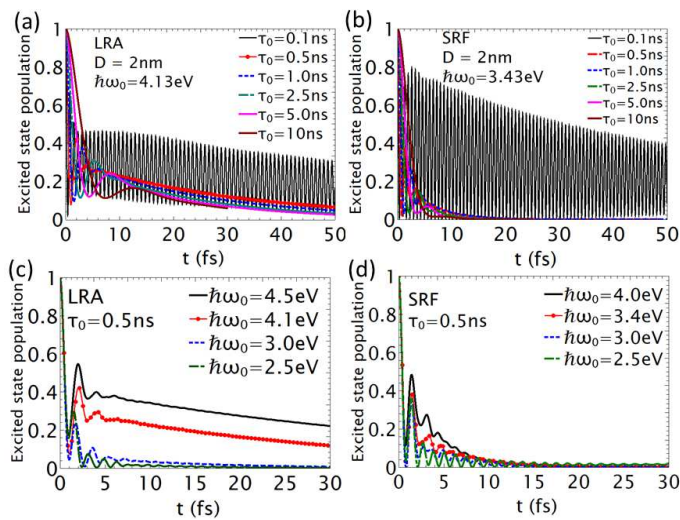}
\caption{Population of the excited state of a two-level QE as a function of
time, within LRA (a,c) and SRF (b,d). The QE is placed in the middle of a MIM
geometry of thickness $D = 2\,$nm. (a,b) The values of $\tau_{0}$ considered are
displayed in the insets. The transition energy of the QE is $\hbar\omega_{0} =
4.13\,$eV (LRA) and $\hbar \omega_{0} = 3.43\,$eV (SRF). (c,d) The values of
$\hbar\omega_{0}$ considered are shown in the insets, for a QE with
$\tau_{0} = 0.5\,$ns.}\label{fig:05}
\end{figure}

In Fig.~\ref{fig:05} we present the excited-state population dynamics of a QE
placed in the middle of a MIM cavity having an insulator (air) layer thickness of
$D = 2\,$nm when different free-space relaxation rates $\tau_{0}$ are considered.
Choosing the associated QE transition energy matching the highest value shown in
the Purcell factor spectra of Fig.~\ref{fig:03}, LRA and SRF models are contrasted
in Figs.~\ref{fig:05}(a,b) for $\hbar \omega_{0} = 4.13\,$eV and $\hbar \omega_{0}
= 3.43\,$eV, respectively. In both panels we observe strong and rapid Rabi oscillations;
as the value of the free-space lifetime decreases from $\tau_{0}=10$ to $0.1\,$ns, the
QE/MIM coupling increases, the oscillation period decreases, and non-Markovian effects
become stronger, as anticipated. We also observe that the population oscillations are
denser in the SRF case; the reason is that what is considered is the full Purcell-factor
spectrum which, compared to the LRA case is broader, and the contribution to the bandgap
region is also substantial, thus leading to a higher QE/MIM cavity interaction.
 The different values of $\tau$ are connected with the different possible
natural or artificial QEs available for experimental manipulation\cite{Bricks2018,Koperski2020,Epstein2020}.

Finally, in Figs.~\ref{fig:05}(c,d) we explore the influence of the QE transition
energy, $\hbar \omega_{0}$, on the population dynamics, taking a QE free-space
lifetime of $\tau_{0} = 0.5\,$ns. Different values of the QE transition energy are
considered, including those matching the SPP energies indicated in the Purcell factor
spectrum of Fig.~\ref{fig:03}. Fig.~\ref{fig:05}(c) shows the LRA calculation, where
we observe that, after a few oscillations, the population dynamics remains partially
trapped---although, eventually, it fully relaxes at later times.
 As the QE transition energy decreases, moving away from the SPP
resonance energy associated with the highest Purcell factor enhancement, the partially
transient trapped population steadily decreases. In the SRF case, we also observe a
highly oscillatory behavior of the excited state population density, a sign that the
non-Markovian signature is retained. Although the partially excited population trapping
is smaller, strong oscillations are still present, with reduced period, indicating that the
overall QE--MIM interaction is higher when the quantum aspects of the metal response are
taken into account.

{\it Summary} --- We theoretically explored the emission properties of a QE placed
in proximity to IM, IMI, and MIM geometries in the weak- and strong-coupling regimes.
Focusing on Na as a plasmonic metal whose nanoscopic optical response is dominated by
electron spill-out, we adopted the quantum-informed SRF approach, in which the Feibelman
$d$ parameters for the centroid of induced charge incorporate first-principles calculations.
Although quantum effects introduce additional losses in the metal, the QE/MIM cavity system
is found to operate in the strong-coupling regime, manifested as a Rabi splitting in the
emission spectrum and through Rabi oscillations in the relaxation dynamics of the QE excited
state, with a highly non-Markovian character. SRF bridges, therefore, \emph{ab initio}
approaches with classical EM calculations in a simple and elegant way, suitable not
only for classical but also for quantum-optical dynamic studies at the nanoscale.

\begin{acknowledgments}{\it Acknowledgments} ---
V.K. research was supported by JSPS KAKENHI Grant Number JP21K13868.
J.~D.~C. is a Sapere Aude research leader supported by Independent Research Fund Denmark (grant No. 0165-00051B).
N.~A.~M. is a VILLUM Investigator supported by VILLUM FONDEN (grant No. 16498).
The Center for Nano Optics is financially supported by the University of Southern Denmark (SDU 2020 funding).
E.P.'s work is co-financed by Greece and the European Union (project code name POLISIMULATOR).
\end{acknowledgments}

\end{document}